\documentclass[fleqn,usenatbib]{mnras}

\usepackage{newtxtext,newtxmath}

\usepackage[T1]{fontenc}

\DeclareRobustCommand{\VAN}[3]{#2}
\let\VANthebibliography\thebibliography
\def\thebibliography{\DeclareRobustCommand{\VAN}[3]{##3}\VANthebibliography}

\usepackage{graphicx}	
\usepackage{amsmath}
\usepackage{natbib}	

\title[UX Ori Stars Eclipses by Disc Perturbations]{UX Ori Stars Eclipses by Large-Scale Disc Perturbations}

\author[S. G. Shulman, V. P. Grinin]{
S. G. Shulman,$^{1}$\thanks{E-mail: sgshulman@gmail.com}
V. P. Grinin,$^{1,2}$
\\
$^{1}$St. Petersburg State University, Universitetskii pr. 28, St. Petersburg, 198504, Russia\\
$^{2}$Pulkovo Astronomical Observatory, Russian Academy of Sciences,
Pulkovskoe sh. 65, St. Petersburg, 196140, Russia
}

\date{Accepted XXX. Received YYY; in original form ZZZ}

\pubyear{2022}

\begin{document}
\label{firstpage}
\pagerange{\pageref{firstpage}--\pageref{lastpage}}
\maketitle

\begin{abstract}
    We simulate the polarized radiative transfer in vicinities of the UX Ori type stars during their minima.
    Our model of an eclipse by an extended disc perturbation generalizes the compact gas-dust cloud eclipse model.
    We apply the radiative transfer method based on enumeration using the directions grid to model the influence of the perturbation extensions along azimuth and radius on the eclipse depth and parameters of the linear polarization.
    We investigate eclipses both for the flared disc and for the disc with a puffing-up in the dust sublimation zone.
    The puffing-up is obtained by adding a disc wind to the model.
    Comparison with a compact cloud eclipse model reveals that the eclipse by a large-scale azimuthally extended perturbation may be significantly deeper and show a greater linear polarization degree.
    We also demonstrate that the perturbation extension together with the disc puffing-up can strongly affect the degree of polarization and colour index of the star during the eclipse.
    The position angle of the linear polarization may also change markedly during and after an eclipse by a large scale perturbation for the model with a puffed-up inner rim.
    Also, in this model, the maximum degree of the linear polarization can be achieved not at the brightness minimum but closer to the end of the eclipse.
    We discuss the modelling results in the context of the photopolarimetric observations of UX Ori stars.
\end{abstract}




\section{Introduction}

Herbig Ae/Be stars are intermediate mass pre-main sequence stars~\citep{Herbig1960}.
Among them, variable UX~Ori stars are distinguished into a separate subclass.
UX~Ori stars demonstrate photopolarimetric variability with deep sporadic fadings up to 2--3 magnitudes and duration from several days to months.
Such fadings are usually accompanied by a rise in the linear polarization degree up to 5--8~per cent.
Selective absorption leads to star colour changes during the eclipse.
When the star fades, it usually turns red first and then turns blue in the deep minimum~\citep{Grinin1991}.
Rarely unusual too deep eclipses with a high polarization degree are observed \citep[e.~g.,][]{Rostopchina2001}.
Sometimes the position angle (PA) of the linear polarization also may change during an eclipse.
It may be explained by the increased contribution of the interstellar polarization with a different PA compared to intrinsic polarization~\citep[see, e.~g.,][]{Shakhovskoi2003}.

\cite{Grinin1988} proposed the eclipse model in which gas-dust clouds obscuring the star are presumed to be sufficiently compact,
and they do not significantly affect the circumstellar disc scattered light.
Consequently, the polarization parameters of the scattered disc radiation do not change during an eclipse.
Thus, the photopolarimetric changes are explained by the increasing contribution of the polarized disc radiation to the system total light, while the star is eclipsed by gas-dust cloud.
The gas-dust cloud can be some kind of a structure in the disc atmosphere, as well as in the disc wind.
Usually the shape of the cloud and its nature are outside the scope of the model.
Hence, this model has only one parameter: the optical depth of the cloud along the line of sight at a certain wavelength $\tau_\lambda$.
As a result, the dependences of the polarization degree and colour indices on the fading level are always the same for a given disc model
(some scatter of values may present due to fluctuations of the scattered radiation and observation errors).
The observed radiation intensity is determined as $I_{obs} = I_* \cdot \exp(-\tau_\lambda) + I_{sc}$,
where $I_*$ is the direct star light intensity and $I_{sc}$ is the intensity of the scattered by the disc radiation.
The observed polarization vector also has a simple dependency on the fading level $\Delta m$: $\overline{P}_{obs}(\Delta m) = \overline{P}_{is} + \overline{P}_{in}\cdot10^{0.4 \Delta m}$.
Here $\overline{P}_{is}$ is the interstellar polarization and $\overline{P}_{in}$ is the intrinsic polarization.
Application of this relation to observations allows one to compose an overdetermined system of equations and determine both components with high accuracy \citep[see, e.~g.,][]{Shakhovskoi2003}.
This approach is not applicable to the cases of prolonged eclipses when the dust screen obscures the inner regions of the disc together with a star, which leads to the intrinsic disc polarization dependency on the eclipse phase.

Thus, the model explains the limitation on the eclipses depth
(the maximum eclipse depth is reached when the direct stellar light is completely absorbed by the cloud),
polarization degree increase and the changes in the colour indices during the eclipses.
Below we call this model a \textquoteleft conservative model\textquoteright\ or a \textquoteleft compact dust cloud model\textquoteright.

Based on the high linear polarization observed at deep brightness minima,
it was suggested that the photometric activity of UX~Ori type stars is caused by the small inclination of circumstellar discs to the line of sight.
The interferometric observations of these stars in the near-infrared wavelengths generally support this hypothesis \citep[see][and references there]{Kreplin2016}.
Exceptions are T~Tauri stars that also demonstrate the extinction events~\citep[see, e.~g.,][]{Ansdell2020}.
Due to their low luminosity, the circumstellar dust can survive in their nearest neighborhood and can penetrate the magnitosphere of the star~\citep{Nagel2020},
which makes extinction events less sensitive to the inclination of their discs to the line of sight.

In various papers, the eclipses were studied within the compact dust cloud model.
In the very first studies, the scattering medium had an elliptical geometry~\citep[e.~g.,][]{Voshchinnikov1989},
later the flared disc model was used~\citep[]{Whitney1992, NattaWhitney2000}.
Often, when modelling eclipses, the dust cloud appeared in the works just as an absorbing screen, and its nature was not considered.
The main focus was on reconciling simulation results with observations.
In general, a good agreement was obtained.

\cite{Natta2001} showed the flared disc model to be unable to explain the near-infrared excesses of Herbig Ae/Be stars.
A new model with a disc puffed up in the inner region (in the dust evaporation zone) was introduced.
Such a model is recognized as the most in line with reality now.
Nevertheless, various possible mechanisms of the puffing-up are considered now~\citep[for a review, see][]{Dullemond2010, Turner2014, Flock2016}.

Originally the heating of the inner disc region by stellar radiation was considered as a cause of the disc puffing-up \citep{Isella2005, Tannirkulam2007}.
Later \cite{Vinkovic2007} shown that radiative heating is not enough to explain infrared excesses and noticed that disc wind may raise dust from the disc surface and lead to disc puffing-up.
\cite{Tambovtseva2008} demonstrated that the dust particles survive in the disc wind, despite the high temperature of the gas.
Thereby the dusty wind could make a significant contribution to the circumstellar extinction, particularly in T Tauri stars.
\cite{Bans2012} used the centrifugally driven winds~\citep[]{Safier1993a, Safier1993b} to model infrared excess of the stars.
They showed that the dusty disc wind may have a significant contribution to the infrared excess.
Besides, the disc wind organizes the radially extended puffed-up region in the inner disc that agrees with the other modelling results \citep{Flock2016} and observations \citep{Lazareff2017}.
Moreover, some theoretical models \citep[e.~g.,][]{Flock2017} produce a vortex in the disc that could possibly create the large-scale disc perturbation presented in our paper.
Therefore, the disc wind can be one of the possible puffing-up reasons.

In our recent paper~\citep[]{Shulman2019a}, we considered the compact dust cloud model of the eclipses for the star with a puffed-up disc.
Following \cite{Bans2012} we also used \cite{Safier1993a, Safier1993b} disc wind models.
Eventually, we obtained new results that did not fit into the disc model without the puffed-up inner boundary:
\begin{enumerate}
\item PA changing by 90\textdegree\ when changing the wavelength.
Some young stars show such changes when switching between optical and near-infrared spectral bands~\citep[]{Pereyra2009}.

\item With a certain combination of the disc wind parameters and the radiation wavelength the polarization degree may not change or change in a narrow range during an eclipse.
\cite{Rostopchina-Shakhovskaja2012} observed such eclipses of the UX~Ori type star WW~Vul.

\item The disc puffing-up may lead to unusual colour changes during the eclipse.
There may not be the star reddening during the fading at all.
The colour indices immediately begin to decrease.
In our model, it is easier to get this behaviour in the blue region of the optical spectrum rather than in the red one.
Such changes of the UX~Ori colour indices were observed from the IUE satellite~\citep[]{Grady1995}.
\end{enumerate}

These results have shown that the dust component of the disc wind could be very important for the intrinsic polarization of young stars.
\cite{Blandford1982} showed that the wind density is strongly connected with the accretion rate.
Therefore the intrinsic polarization of young stars must also depend on the accretion rate.
As a result, the accretion rate fluctuations may lead to polarization variability.
Note that the large scatter of polarization parameters at the same brightness was observed in some of the UX~Ori stars
\citep[e.~g.,][]{Rostopchina2000}.

\begin{figure}
    \center{\includegraphics[width=1\columnwidth]{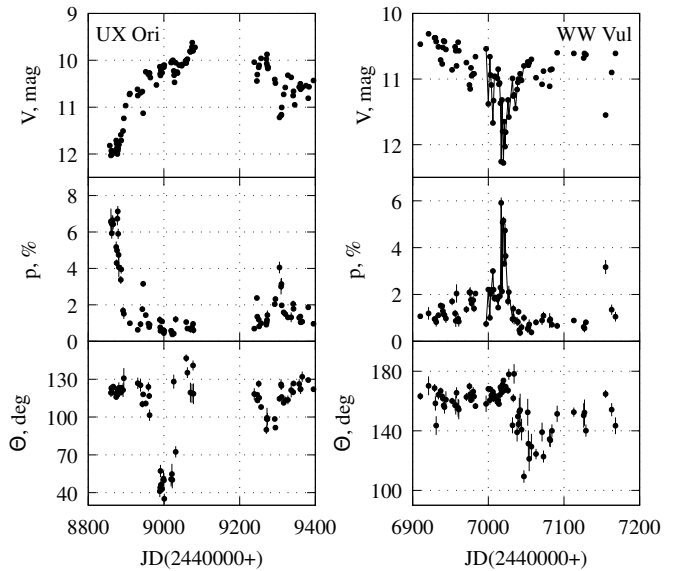}}

    \caption{Magnitude in V band (top panel), polarization degree (medium panel), and PA of the polarization (lower panel)
        during long-lasted deep eclipse of UX Ori~\citep[]{Grininetal1994} and WW~Vul~\citep[]{Grininetal1988}.}
    \label{fig:UXOri}
\end{figure}

The disc puffing-up can explain some unusual polarization parameters changes during the eclipses.
Nonetheless strong polarization parameters changes may also occur shortly after the eclipse.
Significant changes in the linear polarization PA soon after the deep photometric minimum have been observed
in UX~Ori~\citep[]{Grininetal1994} and WW~Vul~\citep[]{Grininetal1988}.
These long eclipses are presented in Fig.~\ref{fig:UXOri}.
They have lasted for a few months.
This is comparable in time to the Keplerian rotation period of the UX~Ori type star disc on the dust evaporation radius.
The compact cloud eclipse model is not able to explain such PA changes,
because the cloud does not affect the disc scattered light.
Similar eclipses have not been observed in other UX Ori type stars yet.

In our second paper~\citep[]{Shulman2019b} we calculated such eclipses for the first time for a flared disc model and a disc with the puffing-up in the inner region.
We used as an extended screen a simple 1-D model of the disc perturbation and received that in the model with a puffed-up disc, the PA changes over time may reach 60\textdegree,
which is comparable to the observational results (Fig~\ref{fig:UXOri}).

In the present paper, we consider the more realistic (for the Keplerian disc) elongated in azimuth disc perturbation and study other features of the eclipses by disc perturbations with different sizes.
We continue to use the dusty disc wind to obtain disc puffing-up since
we have already shown that this approach allows us to explain a number of the observed features of the UX~Ori type star eclipses.
Thus this paper may be viewed as a continuation and a generalization of our previous studies.

We will leave the nature of large-scale disc perturbations outside the scope of this work.
We just note that different physical process may give large disc perturbations.
Among them, there are disc atmosphere vortices~\citep[see, e.~g.,][]{Surville2015, Barge2017},
charged dust grains rising above the disc due to magneto-rotational turbulence~\cite[]{Turner2014},
and accretion and disc wind instability and asymmetry.

\section{Method of calculations}

A lot of different numerical methods may be used to model scattered light intensities and polarization.
Some of the methods are semi-analytical, consider only single scatterings
or have high computational complexity in three-dimensional cases.
The Monte-Carlo method is rather simple in the implementation, does not require analytical integration, does not have a limitation of the number of considered scatterings and works fine in complex three-dimensional studies.
As a result, the Monte-Carlo method is the most popular choice for radiation transfer simulations.

The Monte-Carlo method is based on calculating the photon packets propagating in random directions.
A photon packet propagates in a random direction for a random optical depth.
After that, it is scattered in another random direction.
The procedure may be repeated for a desired number of scatterings.

In this paper we use~\citet{Shulman2018} method to model radiation transfer.
\citet{Shulman2018}~proposed to calculate photon packets propagating into uniformly distributed directions instead of random ones and scatter them after propagating several precalculated optical depth in every direction.
It was shown, that such a technique leads to the same results but significantly faster computationally.
It can be shown that from the mathematical point of view it is the integral calculation in terms of the Riemann sums.
A three-dimensional space partition is a combination of a two-dimensional partition of the unit sphere for propagation direction and a one-dimensional partition by optical depth along each direction.
Within the scope of our task, we considered the star to be a point source of radiation.

The disc mass density distribution, which is required for fast optical depth calculation,
was approximated with an unstructured tetrahedron grid, based on Delaunay's triangulation.
To obtain such a grid \textit{gmsh} program was used~\citep[]{Geuzaine2009}.
This kind of grids allows elements that vary greatly in size.
As a result, we can use small elements in the central region of the disc, where high-density gradients are,
and large ones at the disc periphery with low-density gradients.
Thus, this way we can more closely model more significant and complex areas,
while large simple areas can be modelled relatively quickly.
Moreover, tetrahedron grid-based continuous function interpolation is also continuous,
which is also well for numerical simulation.

In this paper, we limited ourselves to taking into account the contribution of the first four scatterings.
In the appendix~\ref{appendix:multiple}, we consider the contribution of multiple scattering using the model from the section~\ref{sec:res_eclipses} as an example.

\section{Disc Model}

We consider a model of a single early spectral type A star with a flared circumstellar disc.
The star radius is $R_* = 2.9 R_\odot$, and mass is $M_* = 2 M_\odot$.
A disc wind is used to produce a puffing-up in the inner region of the disc, in the dust evaporation zone.
We study the eclipses of the star by a large-scale disc perturbation.

\begin{figure*}
    \center{\includegraphics[width=2\columnwidth]{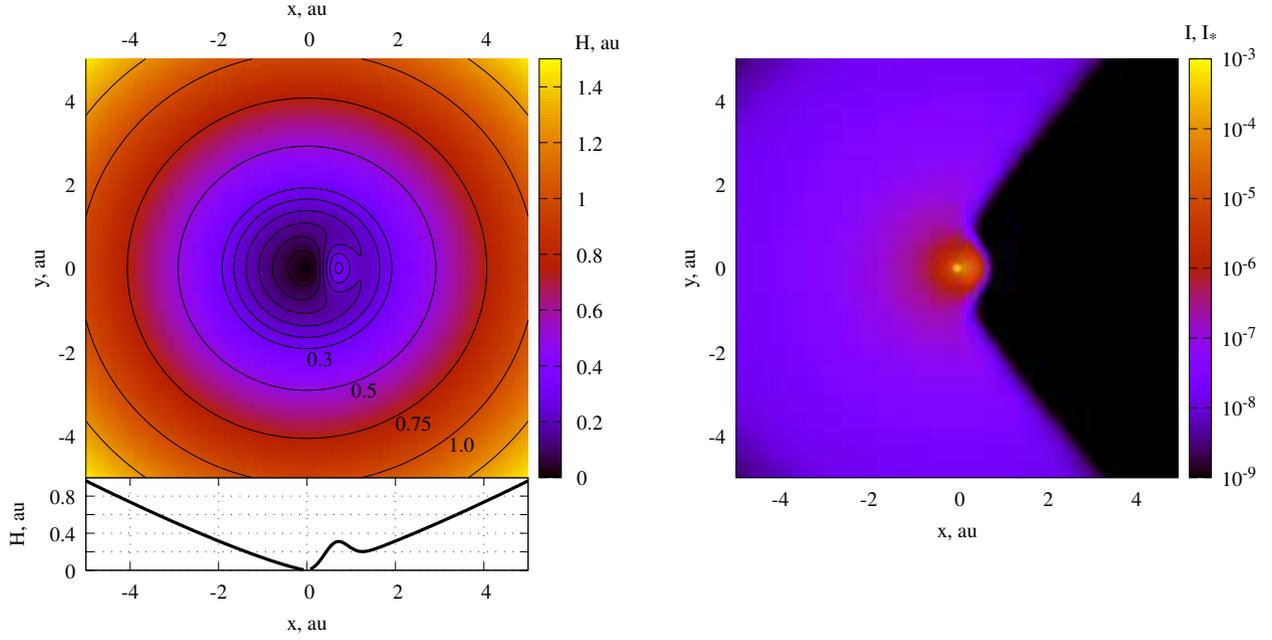}}

    \caption{Left panel: the disc effective optical height is shown at the top.
    The lines connect points with heights equal to $0.05$, $0.1$, $0.15$, $0.2$, $0.25$, $0.3$, $0.5$, $0.75$, $1$, and $1.25$ au.
    Under the heightmap, its cross-section with $y = 0$ is shown.
    Right panel: the image of the disc with a large-scale perturbation observed from the pole.
    The disc radiation intensity is shown in units of the stellar luminosity per 1~au$^2$ .
    To pass to the radiation flux, the intensity presented on the graph should be multiplied by the stellar luminosity and divided by
    the square of the distance to the star expressed in astronomical units.}
    \label{fig:disk}
\end{figure*}

\subsection{Disc}\label{subsec:disc}

Following previous studies \citep[e. g.][]{Shulman2019a, Shulman2019b, Teixeira2009, Robitaille2011} we use the following flared disc form:
\begin{equation}
\rho(x,y,z) = \begin{cases}
                  \rho_0 \left(\frac{R_0}{r}\right)^\alpha \exp\left[-\frac{1}{2}\left(\frac{z}{h(r)}\right)^2\right],       & R_{in} \le r \le R_{out}, \\
                  0,& \text{otherwise}.          \end{cases}
\end{equation}
where $r = \sqrt{x^2 + y^2}$ is the radius in the disc plane,
\begin{equation}\label{eq:disc_h}
 h(r)=h_0\left(\frac{r}{R_0}\right)^\beta
\end{equation}
is the scale height.
Other values are disc model parameters.

In this paper we use the same values as in \cite{NattaWhitney2000} and \cite{Shulman2019a}:
$\alpha = 2.79$ is the radial density exponent,
$\beta = 1.29$ is the flaring power,
$R_{in} = 4R_*$ and $R_{out} = 100$ au are the inner and outer disc radii respectevly.
The disc thickness at radius $R_0 = R_*$ is taken to be $h_0 = 0.008R_*$.
The density normalization constant $\rho_0$ is obtained from the disc mass, which is $0.1 M_\odot$.

\subsection{Disc perturbation}

We use a disc perturbation model based on two Gaussian functions determining the hump shape by distance from the star and azimuth.
For brevity, below we call this large-scale disc perturbations \textquoteleft hump\textquoteright.
This shape resembles the vortex structures (cyclones and anticyclones) predicted by the gas-dynamic models of protoplanetary discs
\cite[see, e.~g.,][]{Godon2000,Wolf2002}.
Close geometry of the perturbation was previously used in \cite{OSullivan2005} for AA~Tau photopolarimetry modelling.
In our previous work~\cite[]{Shulman2019b} we used a simpler hump model, determining its shape only by distance from the hump centre.

Below the hump is described in the polar coordinate ($r = \sqrt{x^2 + y^2}$, $\phi = \arctan(y/x)$) system with the equation:
\begin{equation}\label{eq:hump}
h_{\mathrm{hump}}(\phi, r) = h_{\mathrm{hump0}}
\exp\left[-\frac{1}{2}\left( \frac{\phi - \phi_0}{\Delta \phi}\right)^2\right]
\exp\left[ -\frac{1}{2}\left( \frac{r - r_0}{\Delta r}\right)^2\right],
\end{equation}
here $h_{\mathrm{hump0}}$ is a relative hump height in the units of $h$,
$\phi_0$ and $r_0$ are polar coordinates of the hump centre.
$\phi_0$ and the longitude of the observer $\phi_{obs}$ together determine the relative position of the hump and the observer.
As a result, we can always assume $\phi_0 = 0$ and change only $\phi_{obs}$.
$\Delta \phi$ and $\Delta r$ characterize the hump extension along azimuth and radius.

For the height of the disc with a hump we use the following equation
\begin{equation}\label{eq:h_new}
h(\phi, r) =h_0\left(\frac{r}{R_0}\right)^\beta \left(1 + h_{\mathrm{hump}}(\phi, r) \right).
\end{equation}
instead of~\eqref{eq:disc_h}.
Thus, the hump increases the disc height and may lead to star eclipses.

Fig.~\ref{fig:disk} shows the effective optical height $H$ and the image of the disc with a hump observed from the pole.
By the effective optical height, we mean the height at which the optical thickness of the disc,
when approaching its plane, reaches 1.
Usually, the hydrodynamic height $h$ is about 4--5 times smaller than the optical height $H$.
The hump parameters are: $h_{\mathrm{hump0}} = 3$, $\Delta \phi = 0.6$ and $\Delta r = 0.3$ au.

\subsection{Disc wind}\label{subsec:disc_wind}

In the current paper, the disc puffing-up is produced by a disc wind.
We continue to use the disc wind models from~\cite{Safier1993a, Safier1993b}, as in our previous two.
\cite{Safier1993a, Safier1993b} presented approximations for several wind models which lead to various puffing-ups.
For the detailed discussion we selected model C\@.

Three parameters define the radially self-similar wind solution.
The first parameter is $\kappa_w$, the mass loading, which is equal to the normalized mass-to-magnetic flux ratio.
The second parameter is $\lambda_w$, the normalized total specific angular momentum.
$\lambda_w$ characterizes the efficiency of the angular momentum transport of the wind.
The third one, $\xi_0' \equiv \tan \theta_0$, measures the initial inclination of the magnetic field lines.
Where $\theta_0$ is the angle between the poloidal field component and the disc normal at the disc surface.
For model C the parameters have the the following values: $\kappa_w = 0.01$, $\lambda_w = 75.43$, and $\xi_0' = 1.73$.

The disc wind density is described by the expression

\begin{equation}
    \rho_{\mathrm{wind}}(r, z) = \rho_{\mathrm{wind0}} \left( \frac{r}{r_0}\right)^{-3/2}\eta(z/r).
\end{equation}

Here, the function $\eta(z/r)$ is derived from the solution to the gas-dynamic equations.
\cite{Safier1993b} provides approximations for this function for several disc wind models.
$\rho_0$ is the wind density in g$\cdot$cm$^{-3}$ on the disc surface at distance $1$~au from the star.
It can be expressed through the following parameters: 
the mass outflow rate in solar masses per year $\dot{M}_{out}$,
the stellar mass in solar masses $M_*$,
the inner and outer radii of the wind formation region $r_{\min}$ and $r_{\max}$,
the ratio of the vertical speed to the Keplerian speed at the disc surface $\psi_0$, and
the dimensionless height from which the wind begins $h_{\mathrm{wind0}}$:

\begin{equation}\begin{split}
                    \rho_{\mathrm{wind0}} &= 1.064\times 10^{-15}\left( \frac{\dot{M}_{out}}{10^{-7}\textrm{M}_\odot year^{-1}}\right)
                    \left(\frac{M_*}{0.5 \textrm{M}_\odot} \right)^{-0.5} \times \\
                    & \times \frac{1}{\ln\left( r_{\max}/r_{\min}\right)\psi_0 \left(1-h_{\mathrm{wind0}}\xi_0'\right)}.
\end{split}\end{equation}

The approximation for $\psi_0$ is also available in~\cite{Safier1993b}.
The density is not very sensitive to $r_{\min}$ and $r_{\max}$ radii,
so we can use $r_{\max} / r_{\min} \sim 20$~\citep[]{Bans2012}.
\cite{Mendigutia2011} and \cite{Donehew2011} estimate the accretion rate in Herbig Ae/Be stars from $10^{-6}$ to $10^{-9} M_\odot$ yr$^{-1}$.
It is usually estimated that the mass outflow rate is one order less than the accretion rate.
\cite{Bans2012} accept the mass outflow rate as 1--5~per cent of the accretion rate.
Finally, $h_{\mathrm{wind0}}$ is defined from the disc geometry.

As a result, the mass outflow rate and the disc wind model define the disc wind for the considered star with the flared disc.
The dependency on the mass outflow rate is considered in the paper.
Different wind models are compared in the appendix~\ref{appendix:disk_wind}.

\medskip

When we consider the disc with a hump and the disc wind together,
we apply the hump to the disc and use $\max [\rho(x,y,z), \rho_{\mathrm{wind}}(r, z)]$ as the resulting density of the puffed-up disc with the hump.

\begin{table}
	\centering
	\caption{The dust properties}
	\label{tab:dust_properties}
	\begin{tabular}{lcccc}
		\hline
		\ & $\kappa$ cm$^2$ g$^{-1}$ & $\omega$ & $p_l$ & $g$\\
		\hline
		U       & $330$         & $0.525$       & $0.33$    & $0.53$ \\
		B       & $273$         & $0.53$         & $0.37$    & $0.53$ \\
		V       & $225$         & $0.545$       & $0.38$    & $0.48$ \\
        R       & $183$         & $0.54$         & $0.43$    & $0.43$ \\
		I        & $133$         & $0.525$       & $0.5$      & $0.38$ \\
		\hline
	\end{tabular}
\end{table}

\subsection{Dust properties}

For simplicity, the dust properties were assumed to be constant for the entire model (disc, disc wind, and disc perturbation)
and similar to those of grains in the interstellar medium.
The gas to dust ratio in the disc and the disc wind is also considered the same.
It is equal to 100:1.

For the dust scattering computation, we used the Henyey-Greenstein phase function with the approximations for the polarization functions \citep[]{White1979}.
The absorption coefficient $\kappa$, the single-scattering albedo $\omega$,
and the peak linear polarization $p_l$ were taken from \citet{NattaWhitney2000}.

The phase function asymmetry parameter $g$ was taken from \citet{Kim1994}.
We neglect circular polarization and assume the peak circular polarization equal to $0$.
Following \citet{White1979} the skew factor was taken equal to $1$.

Table~\ref{tab:dust_properties} presents the dust properties for five photometric bands.
The absorption coefficients in Table~\ref{tab:dust_properties} are presented, taking the gas to dust ratio into account.

\section{Results for deepest eclipse points}

In this section, we consider the eclipses by large scale perturbations and compare them with the eclipses by a compact dust cloud.
In case of the compact cloud eclipse we vary the only parameter of the eclipse, the optical depth of the cloud along the line of sight, from 0 to infinity.
For each hump, we consider the deepest point of the eclipse, when the hump centre is between the star and the observer ($\phi_{obs} = 0$).
After that, we examine colour indices for different eclipses and briefly discuss the eclipses of the star with a different disc model.

To eclipse a star with a disc hump, the observer must look along the edge of the disc.
Therefore, a significantly smaller range of values is available for the inclination angle $i$ between the observer and the disc axis.
If this angle is too large, we will see the system from the edge, and the disc will cover the star.
On the contrary, if $i$ is too small, the disc hump will not be able to intersect the line of sight.
For the disc model from the section~\ref{subsec:disc}, we take the angle between the direction to the observer and the disc axis equal to 65\textdegree.
This makes it possible to obtain eclipses by a disc hump with not too high relative height (from 2 to 3).

In our work, the disc perturbation is described with four parameters.
For simplicity, we will consider these dependencies separately.
The relative hump height ($h_{\mathrm{hump0}}$) and the hump extension along azimuth ($\Delta\phi$) determine which parts of the disc will be eclipsed.
Together with the hump extension along the radius ($\Delta r$) the relative hump height also defines the optical depth along the line of sight.
The polar radius of the hump centre ($r_0$) affects whether the side regions farther from the star are eclipsed.

\subsection{Dependence on the hump relative height and the extension along azimuth}\label{sec:res_eclipses}

\begin{figure}
    \begin{center}\includegraphics[width=1\columnwidth]{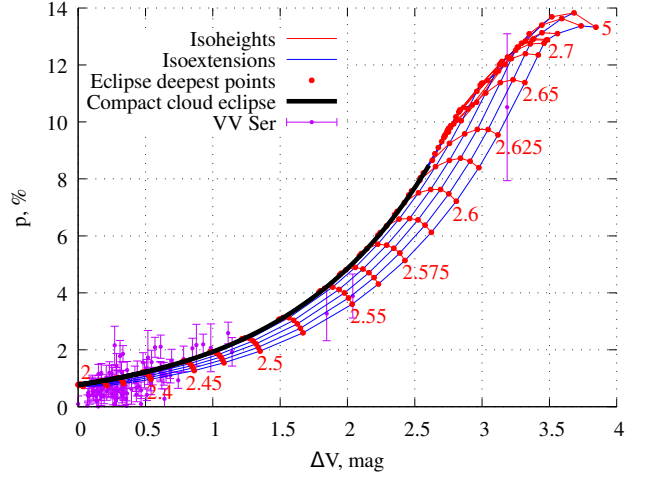}\end{center}
    \caption{The dots present magnitude changes ($\Delta V$) and degrees of polarization ($p$) in the deepest points of the eclipses by humps of different extensions along azimuth and heights.
    The compact cloud eclipse is shown with a thick line.
    Isoheights and isoextensions are plotted with thin lines.
    The heights ($h_{\mathrm{hump0}}$) are labeled next to the isoheights.
    The angle between the direction to the observer and the disc axis is 65\textdegree.
    The unusual deep minimum of VV~Ser is displayed with dots with error bars.}
    \label{fig:eclipse_nw}
\end{figure}

Here we start from dependency on the relative hump height and the hump extension along the azimuth.
An increase in the relative hump height leads to deeper eclipses,
while the change in hump extension along the azimuth gives the maximum variation in polarization depending on magnitude changes.
Fig.~\ref{fig:eclipse_nw} demonstrates points corresponding to maximum magnitude changes ($\Delta V$) and degrees of polarization ($p$)
in the minima of the eclipses with various humps.
For all humps on this figure $r_0 = 0.5$ au and $\Delta r = 0.3$ au\@.
$\Delta\phi$ was changed from $0.03$ to $0.6$.
$h_{\mathrm{hump0}}$ was varied from $0$ to $5$.
\citet{Rostopchina2001} observed an unusual deep minimum of VV~Ser.
It is plotted in Fig.~\ref{fig:eclipse_nw} to demonstrate an eclipse, which may be deeper than a compact cloud model predicts.
We present the intrinsic polarization of VV~Ser, subtracting the interstellar polarization from the observed one.

We assume that the disc perturbation moves with the disc around the star.
To have significant distinguishes from the compact cloud model, it is required to eclipse the inner regions of the disc too.
The duration of eclipses usually ranges from several days to months.
Therefore, we chose $r_0$ value close to the inner radius of the disc to avoid very long eclipses and make it easier to eclipse the disc inner regions.
We also simulated eclipses with different $r_0$ values and obtained similar main results.
$\Delta r$ is limited by the $r_0$: it seems natural that the perturbation has a larger extension in azimuth than in radius.

It should be noted that the eclipse depth varies greatly with small changes in the hump relative height.
At first, the hump height is not enough for any noticeable star eclipse.
Then, with small changes in the hump height, the star quickly weakens.
Finally, the star is completely eclipsed, and, therefore, a further increase in the hump height leads only to
a weakening of the brightness of some areas of the disc.
As a result, the depth of the eclipse increases slightly.
Thus, most of the results in Fig.~\ref{fig:eclipse_nw} were obtained with hump heights from $2.3$ to $2.8$.

Additionally, we show three line types in Fig.~\ref{fig:eclipse_nw}:
the thick line is the graph of the compact cloud eclipse,
thin lines connect points with equal $h_{\mathrm{hump0}}$ and $\Delta \phi$.
We call these lines isoheights and isoextensions respectively.

In general, isoextensions go along the compact cloud eclipse track (see Fig.~\ref{fig:eclipse_nw} for the illustration).
With higher humps, we obtain deeper minima with higher polarization degrees.
The maximum eclipse depth and the polarization degree in the minimum may be significantly greater compared to the compact cloud model.
We achieve $\Delta V$ about $3.7$~mag instead of $2.6$~mag and polarization degree about 14~per cent instead of 8.5~per cent.
Deeper minima are achieved since the inner areas of the disc are also eclipsed together with a star.
The large changes in the polarization degree are associated with different orientations of the scattered light polarization.
Behind the star, the scattered disc radiation is polarized along the disc plane while
the lateral disc parts polarization is perpendicular to the disc plane.
A geometrically thin disc is polarized perpendicular to its plane,
the disc humps eclipse the region behind the star stronger than the lateral areas.
Hence, the total polarization degree of the system increases.
We can see in Fig.~\ref{fig:eclipse_nw} the change in the nature of the dependence of $p$ on $\Delta V$ near the minimum:
the polarization degree almost stops growing and may start to decrease.
This effect is explained by the total star eclipse, due to which the contribution of the disc to the system radiation can not increase anymore.
As a result, all changes occur only because the hump unevenly eclipses different parts of the disc with different linear polarization orientations.

Isoheights go across the compact cloud track.
With narrow humps, we obtain results close to the compact cloud eclipse.
With a wide hump, we get deeper eclipses since the star fading is the same but large areas of the disc are also eclipsed.
The polarization degree changes in a complex way depending on the disc and hump geometry.
In the case of a low hump, the polarization decreases due to lateral disc areas eclipse.
When the disc height is large enough for significant magnitude changes,
with an increase in the hump extension, the polarization first increases and then decreases.
Usually, at the same fading level the polarization in the model with a large scale hump is lower than in the conservative model,
but sometimes it may be slightly higher.

\begin{figure}
    \center{\includegraphics[width=1\columnwidth]{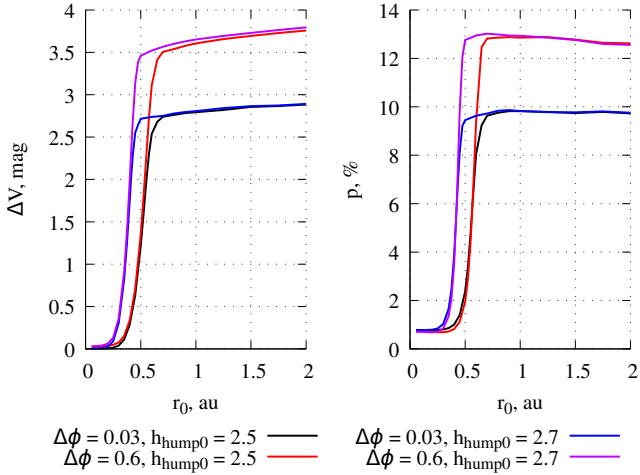}}

    \caption{Magnitude changes (on the left panel) and polarization degree (on the right panel) dependencies on the hump centre radius
    for four hump models. $\Delta \phi$ and $h_{\mathrm{hump0}}$ are listed under the plots. $\Delta r = 0.3$ au for all the humps.}
    \label{fig:r0}
\end{figure}

\subsection{Dependence on the polar radius of the hump centre}

The polar radius of the hump centre ($r_0$) affects the eclipse in two ways.
Firstly, with larger $r_0$ and the same hump extension along the azimuth, the hump can eclipse wider areas of the disc.
So the eclipse depth will increase.
The eclipsed lateral disc regions are polarized orthogonal to the disc plane.
Therefore, when considering a thin disc, the polarization degree will slightly decrease.

Secondly, since we are considering a flared disc,
increasing $r_0$ results in a larger effective hump height at a fixed relative height of $h_{\mathrm{hump0}}$.
At first, the same $h_{\mathrm{hump0}}$ is not enough to eclipse the star, but with larger $r_0$ the star can be completely eclipsed.
In Fig.~\ref{fig:r0} the magnitude changes and polarization degree dependencies on the hump centre radius are shown for four hump models.
All other hump parameters are fixed for each model.

Different $r_0$ leads to small changes in the linear polarization degree corresponding to the fixed eclipse depth compared to the azimuthal extension.

\subsection{Dependence on the hump radial extension}

\begin{figure}
    \center{\includegraphics[width=1\columnwidth]{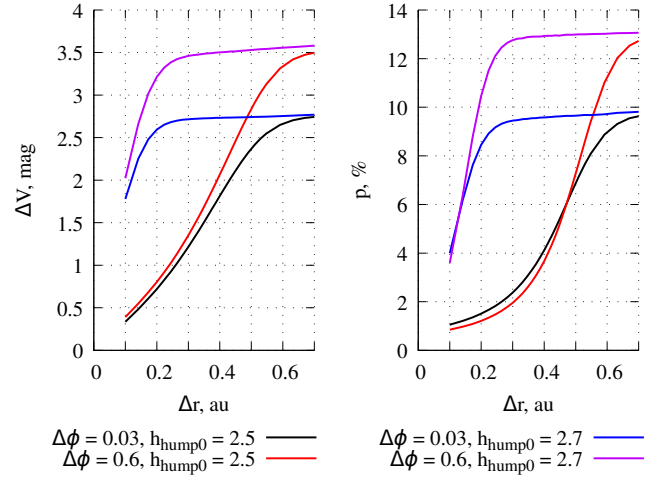}}

    \caption{Magnitude changes (on the left panel) and polarization degree (on the right panel) dependencies on the hump radial extension ($\Delta r$)
    for four hump models. $\Delta \phi$ and $h_{\mathrm{hump0}}$ are listed under the plots. $r_0 = 0.5$ au for all the humps.}
    \label{fig:dr}
\end{figure}

The hump radial extension $\Delta r$ has even less impact on the eclipse.
The dependence on it turns out to be simpler than the dependence on the $r_0$.
In Fig.~\ref{fig:dr} we present the magnitude changes and polarization degree dependencies on the radial extension for four hump models.
A hump with a large radial extension stronger eclipses the star and the disc,
which leads to deeper minima with a large polarization degree.
Gradually, the radial extension becomes large enough so its further increase leads only to a slight increase in the eclipse depth and small changes in the polarization degree.

\medskip

The dependencies on the hump centre radius and the radial extension are not so convenient for increasing the eclipse depth as the dependence on the relative hump height.
Consequently, in the following sections, we will use Fig.~\ref{fig:eclipse_nw} analogues with isoheights
and azimuthal isoextensions to discuss the large-scale disc hump influence on the star eclipses.

\subsection{Disc wind influence}

The numerical modelling showed that the disc wind influence on the eclipses by large scale humps
is very similar to the case of the compact cloud eclipse~\citep{Shulman2019a}.
In this section, we will present a brief demonstration of the wind influence on the eclipses.
It generalizes our previous results to a more complex model of the eclipsing screen.

\begin{figure}
    \begin{center}\includegraphics[width=1\columnwidth]{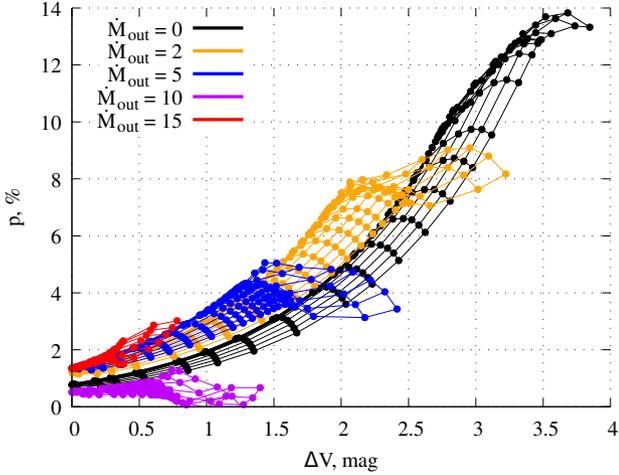}\end{center}
    \caption{The same as in Fig.~\ref{fig:eclipse_nw} for the disc with the disc wind.
    Disc wind model C is used.
    The mass outflow rates are signed in the units of $10^{-9}$ $M_\odot$ year$^{-1}$. }
    \label{fig:wind_mout}
\end{figure}

In Fig.~\ref{fig:wind_mout} we show eclipses obtained with C wind model and different mass outflow rates.
The observer position and the parameters of all humps are the same as in Fig.~\ref{fig:eclipse_nw}.
The considered mass outflow rates are $0$, $2\cdot 10^{-9}$, $5\cdot 10^{-9}$, $10^{-8}$, and $1.5\cdot 10^{-8}$ $M_\odot$ year$^{-1}$.
The results with $0$ mass outflow rate are the same as in Fig.~\ref{fig:eclipse_nw}.

With an increase in the mass outflow rate,
the disc wind scatters additional stellar radiation.
As a result, for observers near the pole of the disc, the system becomes brighter.
On the contrary, when the angle between the direction to the observer and the disc plane is small,
the direct stellar light is absorbed in the wind during the propagation to the observer.
Consequently, the contribution of the star and disc inner regions to the total radiation of the system decreases.
This leads to the fact that eclipses become less deep.
Linear polarization degree changes are more complicated:
when the wind density is low, the wind increases the polarization degree on the same level fadings;
with higher wind density the polarization degree decreases to zero;
finally, with a further increase in the mass outflow rate, the degree of polarization begins to grow.
The nature of this behaviour is the fact that the wind thickens the disc,
therefore the orientation of the linear polarization changes.
For a thin disc with a light wind, it is perpendicular to the disc plane.
Otherwise, for a thick disc with a dense wind, the linear polarization is parallel to the disc plane.
Some isoextensions on the plot reach zero polarization.
On such isoextensions there are eclipses with different linear polarization orientations.
When the disc is in the borderline state between thick and thin ones,
large scale perturbation, eclipsing various disc areas, may have a major influence on the linear polarization orientation.

\begin{figure}
    \begin{center}\includegraphics[width=1\columnwidth]{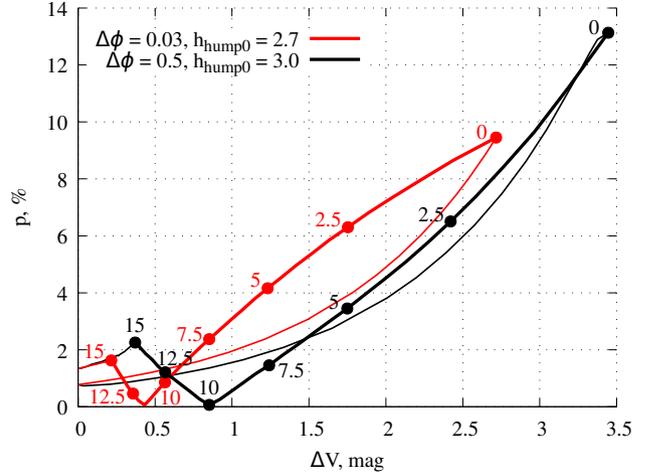}\end{center}
    \caption{Degree of linear polarization in the V band versus change in magnitude caused by a large-scale perturbation eclipse.
    The disc with the disc wind of model C is considered.
        $r_0 = 0.5$ au and $\Delta r = 0.3$ au for the both hump models.
        The thick lines connect points with a constant hump shape and an increasing mass outflow rate.
        The mass outflow rate defining the wind intensity was varied from $0$ to $1.5 \times 10^{-8}$ $M_\odot$ per year.
        The dots on the lines correspond to the mass outflow rates of $0$, $2.5$, $5$, $7.5$, $10$, $12.5$, and $15 \times 10^{-9}$ $M_\odot$ per year.
        The mass outflow rate is indicated next to each point.
        The break of the thick line (dot with $p=0$) corresponds to the change in the PA of the linear polarization by 90\textdegree.
        Thin lines show isoextensions for the minimum and maximum mass outflow rates and fixed other hump parameters.}
    \label{fig:wind_mout_lines}
\end{figure}

Fig.~\ref{fig:wind_mout_lines} shows in detail the changes in the eclipse parameters with a change in the mass outflow rate and, therefore, the wind density.
This figure is very similar to the graphs from our paper~\citet{Shulman2019a},
only we replaced the graphs for compact cloud eclipses by isoextensions, as the closest to them in behaviour and meaning.
Only isoextensions for the minimum and maximum mass outflow rate are shown in the figure.
Intermediate ones run along with them to the constant hump shape line.
We can conclude that, in general, the disc wind has the same effect on eclipse by a large-scale disc perturbation as in the conservative model:
it makes possible an eclipse without an increase in the linear polarization degree and
the PA changing by 90\textdegree\ when changing the wavelength.
Additionally, it increases the possible extension of the linear polarization degree at a fixed fading.

When the disc wind density increases we can see one more interesting result:
the scatter of polarization parameters at one brightness level strongly increases for deep eclipses.
This is because the star is almost completely hidden from the observer by the disc wind and hump.
As a result, polarization does not increase due to an increase in the contribution of the scattered by the disc radiation.
On the contrary, the most important thing turns out to be which parts of the disc are eclipsed.
So, the humps with different extensions give us eclipses with very different polarization degrees.
In this model, the mass outflow rate when the polarization orientation changes depends not only on
the wavelength, disc model, and the observer position, as in the compact cloud case
but also on the hump shape.

Different disc wind models lead to similar results but at different mass outflow rates.
In particular, when using different models, the rotation of the linear polarization plane by 90\textdegree\
(transformation of the disc from thin to thick) occurs at different mass outflow rates.
We will not discuss these distinctions in detail here.
A brief comparison of the simulation results for three disc wind models is presented in the appendix~\ref{appendix:disk_wind}.

\begin{figure*}
    \begin{center}\includegraphics[width=2\columnwidth]{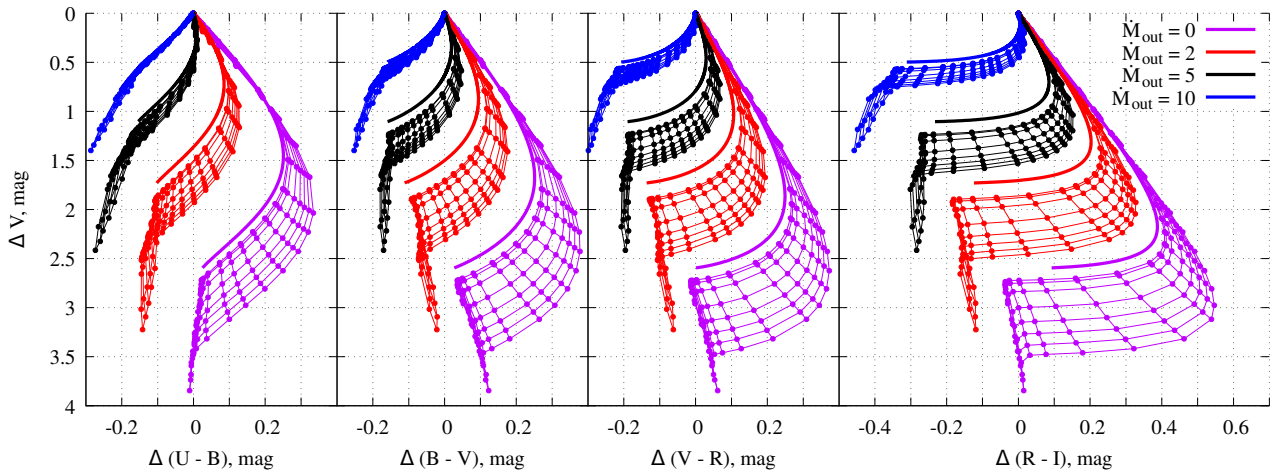}\end{center}
    \caption{Changes in colour indices during eclipses for discs with different wind densities.
    The mass outflow rate in units of $10^{-9}$ $M_\odot$ year$^{-1}$ is signed on the graphs.
    Thick lines show colour indices for the compact cloud model, dots show colour indices for large scale perturbations,
    and thin lines are for isoheights and isoextensions.}
    \label{fig:dUBVRI}
\end{figure*}

\subsection{Colour indices}

Above we considered the eclipses in the V band.
The influence of large-scale disc hump on the eclipse parameters in different spectral bands is approximately the same.
In all bands, we obtain deeper minima with higher polarization degrees.
The large-scale perturbation eclipses reach almost $4$~mag, while the depth of the compact cloud eclipse is about $2.6$~mag.
The linear polarization degree differs from band to band.
The compact cloud eclipse model gives us maximal linear polarization degree equal to 7~per cent in the U band
and 13~per cent in the I band.
The large-scale perturbation model increases maximal polarization degree to 12 and 18~per cent in U and I bands respectively.

In the conservative model, we obtain colour changes due to selective absorption.
The colour indices increase during the star fading and then decrease near the photometric minimum.
This is called \textquoteleft blueing effect\textquoteright.
A large hump also eclipses the inner regions of the disc.
As a result, we obtain larger colour indices on the same fading level.
Moreover, in the deep minimum, the behaviour changes, when the star is fully eclipsed and colours change due to the inner disc regions eclipse.

In Fig.~\ref{fig:dUBVRI} colour indices for the results from Fig.~\ref{fig:wind_mout} are presented.
Only the data with $\dot{M}_{out} = 1.5 \cdot 10^{-8}$ $M_\odot$ year$^{-1}$ is omitted as $\Delta V$ was too low.
The compact cloud model colour indices are shown with the thick lines.
It is interesting, that the eclipse of the inner regions of the disc turns out to be more significant for colour indices than for the polarization degree.
On the figures with the polarization degree versus the magnitude change the results for a hump with a small azimuth extension
almost coincided with the result of the conservative model.
On the contrary, in Fig.~\ref{fig:dUBVRI} there are noticeable gaps between compact cloud model results and
the results of the model with the large-scale perturbation.

We have already obtained earlier in the compact cloud eclipse model
that the disc wind instability increases the possible variation of the colour indices at the same fading level.
Now we can see that the ability to change the size and shape of the absorbing screen also provides such an effect.
Together, they allow explaining very strong scatter in the colour indices during different eclipses of the same star.

\subsection{Different disc models and observer positions}

The above results are presented for only one disc model and one observer position.
Of course, the choice of these parameters also affects the result.

We have already mentioned above that the choice of the angle between the disc axis and the direction towards the observer is strongly limited by the fact
that the star should be clearly visible before an eclipse, and the deep minimum must be achieved at not too large disc height.
A variation of the observer's position in this range changes the eclipse depth and the maximum degree of polarization.
Nonetheless, the hump shape and disc wind influence the eclipse in the same manner.

We performed eclipse simulations for the same disc model, the results of which are presented in Fig.~\ref{fig:eclipse_nw}  for other observer positions.
When the angle between the disc axis and the direction towards the observer is 55\textdegree\ instead of 65\textdegree,
a higher hump is needed to eclipse the star by a large-scale disc perturbation.
Here we have $h_{\mathrm{hump0}}$ from $4$ to $4.5$ instead of from $2.3$ to $2.8$ for 65\textdegree.
In this case, we get an even greater relative increase in the linear polarization degree~--- almost 2 times (from 6.5 to 12.2~per cent) with approximately the same increase in the eclipse depth.
Moreover, at the same fading level the polarization in the model with a large scale hump is slightly higher than in the conservative model.

On the contrary, if the angle between the disc axis and the direction towards the observer is increased from 65\textdegree\ to 68\textdegree,
then a hump with a lower $h_{\mathrm{hump0}}$ will be sufficient (in our case, less than $2.2$), the relative increase in the polarization degree and the eclipse depth will decrease.
The spread of polarization parameters at the same fading level will also be reduced.
At a higher angle, the star will be observed through the disc and eclipsed by it.

Also, there are some minor changes in secondary parameters of eclipses, such as the mass outflow rate, at which the linear polarization orientation rotates.
These changes are of no fundamental importance and can be omitted in the current paper.
Thus, the nature of the phenomena is preserved when the observer position changes, although the numerical results vary.

The parameters of the disc also affect the characteristics of the eclipse.
For the eclipses by large-scale disc perturbations, the disc thickness is the most important parameter.
It determines the possible range of the observer positions and noticeably affects the hump height required to obtain a deep eclipse with the remaining parameters fixed.
Other disc parameters also affect the numerical characteristics of the eclipse: the depth of the eclipse,
dependence of the polarization degree and orientation on the fading rate, colour indices, and others.

All the features of the influence of the hump shape and puffing-up in the dust sublimation zone are preserved.
For example, in the considered model in V band large-scale perturbation increases $\Delta V$ from $2.6$~mag to $3.7$~mag
and the polarization degree from $8.5$ to $14$~per cent compared to the compact cloud model.
We also considered UX Ori disc with parameters from~\citet{Kreplin2016} and the same observer position.
The conservative model gives for this disc a deep minimum with $\Delta V = 3$~mag and polarization degree about $8.3$~per cent.
The large-scale perturbation model allows us to get $\Delta V$ up to $4.4$~mag and polarization degree up to $14$~per cent again.
The changes in the polarization degree turned out to be the same, and $\Delta V$ increased by about 1.4 times.

The puffing-up in the dust sublimation zone can lead to a change in the disc shape.
Instead of a flared disc, we will have a self-shadowed disc~\citep{Dullemond2001}.
In the model with a self-shadowed disc, the impact of the puffing-up will increase even more.
The results for a self-shadowed disc with a certain mass outflow rate determining the puffing-up will be close
to the results for a flared disc with a higher mass outflow rate and a stronger puffing.
This is a result of the fact that the peripheral regions of the self-shadowed disc contribute less to the total system radiation compared to the flared disc ones.

Thus, we checked that the nature of the phenomena is the same for different disc models, but illustrated them in detail for only one.

\section{Eclipse phase dependence}

In the previous section, the closest line to an eclipse was the isoextension when the star's light weakened with an increasing hump height.
We believe that a more realistic scenario for the star fading is not an increase in the hump height,
but the movement of a high hump around the star, in which the hump gradually covers the star.

To simulate the motion of the hump around the star, we used different values of $\phi_{obs}$.
To convert results to time dependency, one should take into account the period of the disc rotation at the $r_0$ distance.
Since in this work, we restrict ourselves to a common study of the phenomenon,
and not to comparison with the specific observations, we will omit this step and present the time dependence in the geometric term of $\phi_{obs}$.

It is worth noting that in our model the hump height $h_{\mathrm{hump0}}$ is considered relative to the flared disc without any puffing-up.
Accordingly, when the puffing-up is significant, the actual hump height is smaller than $h_{\mathrm{hump0}}$.
In the previous sections, the results of our models were weakly sensitive to the maximum considered $h_{\mathrm{hump0}}$ value.
For significant deviations of the PA, which we will study in this section, large disc perturbations are needed.
This is the reason to increase the maximum considered value $h_{\mathrm{hump0}}$ from $5$ to $6$ in this section.

The possibility of varying several parameters at once allows us to consider many models of eclipses.
Nevertheless, we restrict ourselves to a relatively small number of them, allowing us to get an idea of the possible effects.
Below, we will omit the change in $\Delta r$, which mainly affects the optical thickness of the hump,
and $r_0$ since it complicates the result comparison and adds the need to choose a different $h_{\mathrm{hump0}}$ value due to the flared disc shape.
However, larger $r_0$ allows changing wide areas of the disc and enhances some obtained effects.

\subsection{Results for a flared disc}

\begin{figure}
    \begin{center}\includegraphics[width=1\columnwidth]{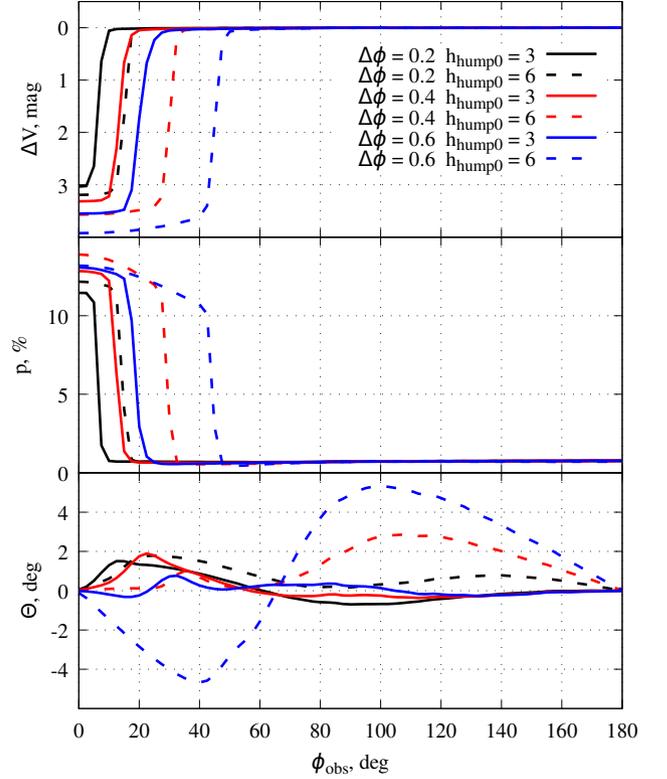}\end{center}
    \caption{Change in magnitude (top panel), degree of polarization (medium panel), and polarization PA (lower panel) in the V band versus longitude of the observer ($\phi_{obs}$).
        For all hump models $\Delta r = 0.3$ au and $r_0 = 0.5$ au.
        $h_{\mathrm{hump0}}$ and $\Delta \phi$ are listed on the plot for each line.
        The results are presented for the flared disc without the puffed-up inner rim.
        The inclination angle $i = 65$\textdegree.}
    \label{fig:time_dependence_w0}
\end{figure}

Fig.~\ref{fig:time_dependence_w0} shows the dependence of the radiation parameters (magnitude changes, the polarization degree and the PA)
on the observer longitude $\phi_{obs}$ for a disc without puffing-up in the dust sublimation zone.
There is nothing unusual about these eclipses: the maximum degree of linear polarization is achieved at the minimum brightness.
With the increasing brightness of the star, the degree of polarization decreases.
The linear polarization PA varies within five degrees in both directions.

The observed changes in the PA after passing through the minimum distinguish the presented eclipses from those in the compact gas-dust cloud model,
but we can hardly consider these changes significant.
The reason for these changes is that the disc hump violates the disc symmetry.
The large-scale perturbation influences the polarized scattered radiation of different disc areas:
the hump can both additionally eclipse some areas of the disc and increase the scattered radiation.
Therefore, its effect on the PA can be different and depends on the geometric properties of the disc, the hump and the observer position.

The observed behaviour weakly depends on the parameters of the hump:
the depth of the eclipse, the maximum linear polarization degree, the duration of the eclipse phases can change.
Minor variations in the PA can also present, but the shape of the eclipse is about the same.

\begin{figure}
    \begin{center}\includegraphics[width=1\columnwidth]{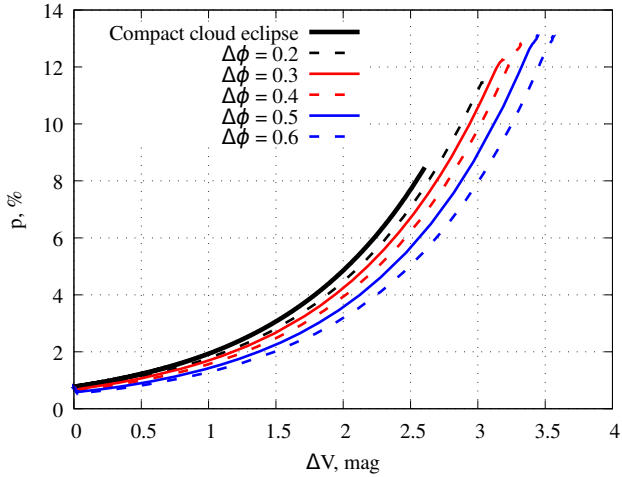}\end{center}
    \caption{Degree of linear polarization in the V band ($p$) versus change in magnitude ($\Delta V$)
        during the eclipse by a moving large-scale disc perturbation.
        The compact cloud eclipse is shown with a thick line for comparison.
        The observer position and the disc model are the same as in Fig.~\ref{fig:time_dependence_w0}.
        For all humps models $h_{\mathrm{hump0}}=3$, $\Delta r = 0.3$ au, and $r_0 = 0.5$ au.
        $\Delta \phi$ is listed on the plot for each line.}
    \label{fig:time_polarization}
\end{figure}

From the results in Fig.~\ref{fig:time_dependence_w0}, it is difficult to assess how the degree of polarization depends on the fading level.
We have shown this dependence in Fig.~\ref{fig:time_polarization} for the five humps with different $\Delta \phi$.
For comparison, the conservative model eclipse is also presented in it.
The lines of eclipses by moving large-scale disc perturbations go along the compact cloud eclipse line and isoextensions in Fig.~\ref{fig:eclipse_nw}.
For each $\Delta \phi$ the resulting graph differs from the corresponding isoextension,
because the hump has a different shape and eclipses the disc differently.

A largely displaced disc hump weakens the polarized scattered disc radiation more strongly than a small hump with the centre exactly on the line of sight.
As a result, the considered eclipses have a lower polarization degree than isospreads on the same fading level,
and differ even more from the conservative model eclipses.
Thus, the shape of the eclipse from Sect.~\ref{sec:res_eclipses} (by a hump with the constant position and an increasing height)
and the shape of the eclipse from the current section (by moving around the star hump with a constant shape) in general turn out to be the same.
However, in the second case, it is possible to achieve even greater parameters scatter on the diagram presenting the linear polarization versus the change in magnitude.

\subsection{Results for a puffed-up disc}

For the disc with a puffing-up in the dust sublimation zone (modelled with the disc wind~\ref{subsec:disc_wind}), the eclipses begin to change noticeably.
Therefore, we considered several models with a constant hump shape and an increase in the mass outflow rate.
As follows from the results presented above, with increasing puffing-up,
the depth of eclipses and the polarization degree at the minimum decrease.
But the differences do not stop there.

\begin{figure}
    \begin{center}\includegraphics[width=1\columnwidth]{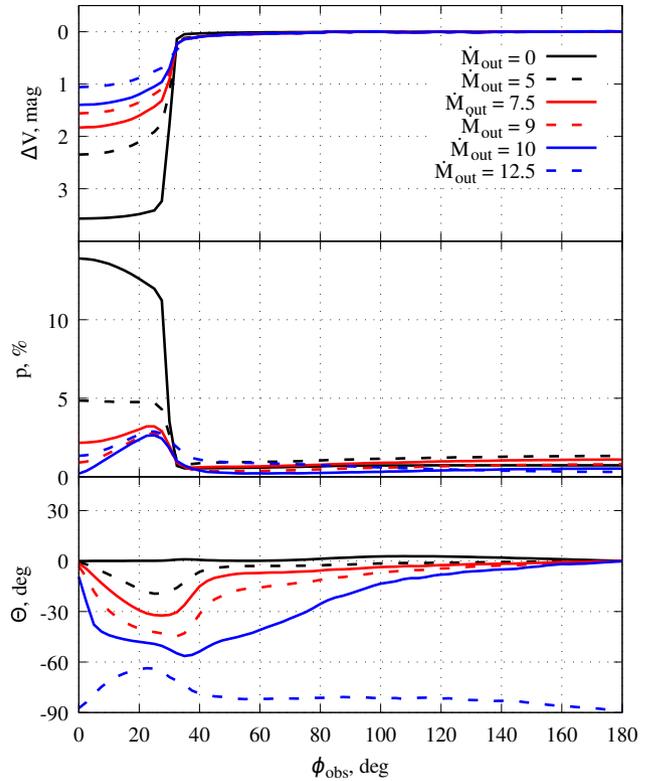}\end{center}
    \caption{The same as in Fig.~\ref{fig:time_dependence_w0} for one hump model ($\Delta r = 0.3$ au, \mbox{$r_0 = 0.5$~au}, $h_{\mathrm{hump0}}=6$ and $\Delta \phi = 0.4$) and a disc with the disc wind.
    The wind model C with different mass outflow rates is used.
    The mass outflow rates is specified in $10^{-9}$ $M_\odot$ year$^{-1}$. }
    \label{fig:time_dependence_dw}
\end{figure}

In Fig.~\ref{fig:time_dependence_dw} we present results for the constant hump shape and an increasing mass outflow rate,
which defines the disc puffing-up.
With an increase in the mass outflow rate, we can pay attention to the following features on the eclipses:
the depth of the eclipse and the polarization degree in the minima decrease;
from a certain mass outflow rate, the maximum degree of linear polarization is achieved not at the minimum, but after it, when $\phi_{obs} \sim 25^\circ$;
the amplitude of the PA deviations increases.

The maximum of the linear polarization degree is shifted from the brightness minimum
because the stellar radiation and scattered disc radiation are strongly absorbed by the extended disc hump and the puffing-up.
In such a situation the degree of polarization depends not only on the intensity ratio of unpolarized stellar radiation to polarized disc radiation (as in the conservative model),
but also on the polarization of the scattered radiation.
The disc with a large-scale perturbation becomes highly asymmetric.
As a result, the scattered light polarization can change noticeably, depending on the geometric arrangement of the observer and the perturbation.
In our case, the Stokes parameter Q weakly changes when the hump eclipses the star.
On the contrary, the Stokes parameter U is zero for the symmetrical geometry (when the hump centre is between the star and the observer)
and grows rapidly with increasing $\phi_{obs}$.
Consequently, the disc scattered light polarization increases noticeably.
There may also be a slight increase in linear polarization degree when the hump and the observer are on opposite sides of the star.

The mass outflow rate and the disc wind model determine the disc puffing-up in the dust sublimation zone.
The greater the mass outflow rate, the thicker the disc becomes.
A thicker disc is polarized perpendicularly to its plane weakly than a thin one.
Hence large-scale disc perturbations lead to stronger changes in the linear polarization PA.
This dependence on the mass outflow rate is retained until the disc becomes thick enough that its scattered light is polarized along the disc plane.
Thereafter, a further increase in the mass outflow rate will cause the amplitude of the PA changes to decrease.
The maximum angle changes are possible for the disc with almost unpolarized radiation.
In Fig.~\ref{fig:time_dependence_dw} the considered model gives deflections of the PA up to $60$\textdegree\ for a thin disc and up to $30$\textdegree\ for a thick disc.
Significant PA deviations are observed both during the eclipse and after it.
In the presented results, the deviations are mainly in one direction.
Before the eclipse, our model predicts symmetrical deviations into opposite direction.
WW~Vul observations~\citep{Grininetal1988} in Fig.~\ref{fig:UXOri} do not demonstrate significant deviations into opposite direction before the eclipse.
The minimum was highly asymmetric.
An asymmetric absorbing screen or a variable mass outflow rate should be considered to simulate it in the future.
This difference between the model and observations indicates the need for further development of the model.
In UX~Ori case, there were no observations during $150$ days before the presented minimum, so we are unable to estimate its symmetry.

\begin{figure}
    \begin{center}\includegraphics[width=1\columnwidth]{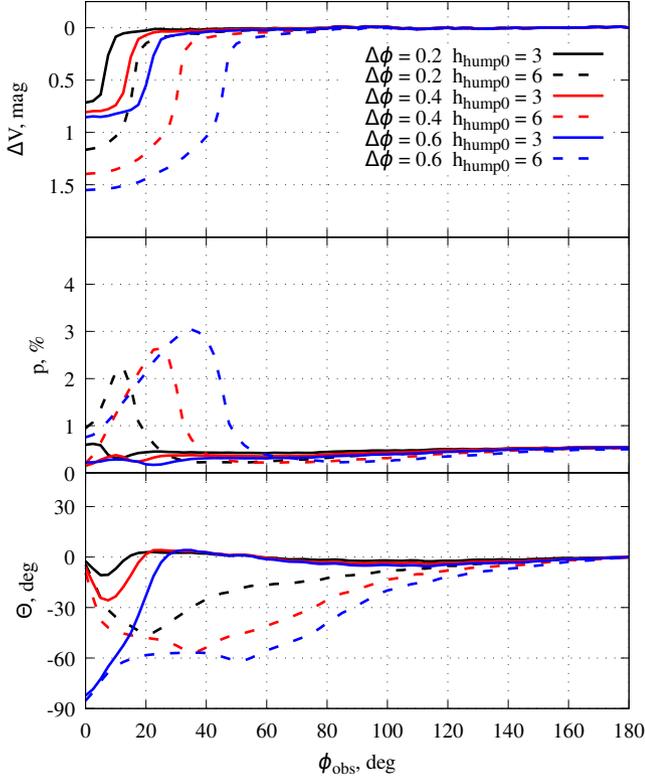}\end{center}
    \caption{The same as in Fig.~\ref{fig:time_dependence_w0} for the same hump models and a disc with the disc wind.
    The wind model C with the mass outflow rate $10^{-8}$ $M_\odot$ year$^{-1}$ is used. }
    \label{fig:time_dependence_w10}
\end{figure}

In order to present the eclipses in more detail in Fig.~\ref{fig:time_dependence_w10} we show eclipses for all hump models from Fig.~\ref{fig:time_dependence_w0} for a disc with the model C disc wind.
The mass outflow rate is $10^{-8}$ $M_\odot$ year$^{-1}$.
With a small hump height, a large deviation of the PA is obtained during an eclipse.
And with a more significant hump, the essential change in the PA can persist for some time after the end of the eclipse.
Moreover, in the brightness minimum, there may be no change in the PA at all or its change by $90$\textdegree.
The last option is observed when the hump has a large azimuthal spread.
This is explained by the fact that the extended large-scale disc perturbation,
located between the star and the observer, increases the effective disc puffing-up for the observer.
Therefore, a thin disc with a slight puffing-up and a large hump can behave like a thick disc with a strong puffing-up.
When the hump shifts around the star, the observer sees a thin disc again.

In our previous paper \cite[]{Shulman2019b}, round hump did not give us 90\textdegree\ PA change during the eclipse.
Herewith the maximum deviation of the PA was achieved after the eclipse, and not during it, as in the above examples.
We have found that even simple models of the disc hump lead to new features of the PA behaviour during the eclipse and after it.
The PA dependence on the observer longitude is rather sensitive to the hump model, model parameters and the puffing-up.
Nevertheless, we can confidently conclude that for a star surrounded by the disc with a puffing-up,
there might be strong changes in the PA.
These changes are possible for stars with thin and thick discs.
During an eclipse, the changes can reach 90\textdegree, and, after the eclipse, they are up to 60\textdegree.
A star surrounded by a disc without puffing-up can hardly show noticeable changes in the PA during the considered eclipses.

\subsection{Results in different spectral bands}

The disc puffing-up optical depth also depends on the optical properties of the dust and, as a consequence, on the wavelength.
In our models, the disc with a puffing-up turns out to be thinner in the I band than in the U band, as a result,
for a thin disc, the PA changes in the I band are less than those in the U band.

A more complex form of the disc hump opens up additional possibilities for changing the PA of linear polarization.
In particular, the hump may be asymmetrical and have forward and backward parts with different spreads.

\begin{figure}
    \begin{center}\includegraphics[width=1\columnwidth]{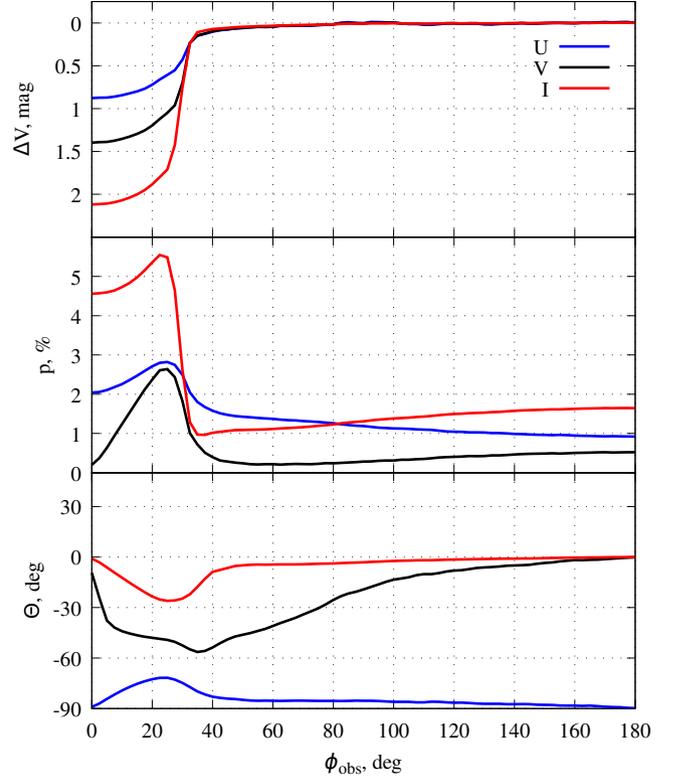}\end{center}
    \caption{The same as in Fig.~\ref{fig:time_dependence_w0} for one scattering geometry model in U, V, and I spectral bands.
    The hump parameters are $\Delta r = 0.3$ au, $r_0 = 0.5$~au, $h_{\mathrm{hump0}}=6$ and $\Delta \phi = 0.4$.
    The wind model C with the mass outflow rate $10^{-8}$ $M_\odot$ year$^{-1}$ is used. }
    \label{fig:time_dependence_UVI}
\end{figure}

Depending on the optical properties of the dust, the puffing-up in the perturbed regions of the disc has a different effect on the shape of the eclipse.
With the higher dust absorption coefficient and the same wind model, we get a more noticeable puffing-up.
As a result, the eclipse will be less deep, and the degree of polarization will be lower.
It is possible that in one spectral band the disc will be thick, while in the other it will be thin.
Then the PA in different spectral bands will be different.
Fig.~\ref{fig:time_dependence_UVI} demonstrates such behavior for three spectral bands.
Consequently, the deviations of the PA during the eclipse will be different:
in one band there will be deviations up to 60\textdegree\ from the orientation perpendicular to the disc plane (V band in Fig.~\ref{fig:time_dependence_UVI}),
and in the other~--- up to 30\textdegree\ from the position along the disc plane (U band in Fig.~\ref{fig:time_dependence_UVI}).
Observation of the PA during a deep eclipse in different spectral bands would be very valuable for studying the inner regions of the disc.

\medskip

We also simulated the changes in magnitude, degree of polarization, and polarization PA versus longitude of the observer for another disc model (UX~Ori disc) and different observer positions.
Since the positional polarization angle is highly dependent on the geometry of the scattering matter,
the results are highly dependent on disc parameters and the observer's position.
However, all of the resulting eclipse features are presented above.

\section{Discussion and Conclusion}

We studied the eclipses of a star by a large-scale disc perturbation and
considered both a model of a geometrically thin flared disc and a model of a disc with a puffing-up in the dust sublimation zone.
The magneto-centrifugally driven dusty disc wind~\citep{Safier1993a,Safier1993b} was used to create the disc puffing-up.
It is rather convenient mechanism of disc puffing-up in the inner regions,
but it is worth remembering that several other physical processes are also leading to disc puffing-up~\citep{Turner2014, Flock2016}.
The study of eclipses of stars surrounded by discs with other puffing-up models is an important further step.
It will make it possible to generalize the results of this work and find whether the obtained effects depend on the shape of the puffing-up and whether the eclipse observations may be used to investigate disc puffing-up shape.
We compared the results obtained in the large-scale perturbation model with the results for the compact gas-dust cloud model~\citep{NattaWhitney2000,Shulman2019a}.

We found that our model with the puffed-up disc allows one to interpret the following observed phenomena:
\begin{enumerate}
    \item Change in the orientation of the polarization plane by 90 degrees when changing the radiation wavelength, that has been detected, e.g., by \citet{Pereyra2009}.
    \item No (or small) increase in the polarization degree during an eclipse observed for WW~Vul by \citet{Rostopchina-Shakhovskaja2012}.
    \item The colour changes during the eclipse without the star reddening during the fading found for UX~Ori by \citet{Grady1995}.
    \item An increase in the parameters scatter (degree of polarization and colour indices) corresponding to the same fading level.
    A significant scatter of these parameters has been observed for several stars~\citep[e.~g.,][]{Rostopchina2000} and can be explained by various reasons.
    Our model suggests one more possible explanation.
\end{enumerate}

When we consider the model of an eclipse by a large-scale disc perturbation with a small extension,
the eclipse will be very similar to an eclipse by a compact dust cloud.
However, regardless of the disc puffing-up in the dust sublimation zone,
the model of an eclipse by a large-scale disc perturbation has important differences from the compact gas-dust cloud eclipse model.

Firstly, in the large-scale perturbation model we obtain a significant restriction on the observer position:
the angle between the direction towards the observer and the disc plane can not be large so that the line of sight passes through the disc perturbation.
Interferometric observations show that UX~Ori stars are indeed observed at a low angle to the disc plane \citep[e.~g.,][]{Kreplin2016}.
Thus, this model feature does not contradict the observations.

Secondly, the absorbing extended screen eclipses not only the star but also the central regions of the disc.
As a result, we get deeper minima with a higher polarization degree.
The minimum depth can increase by about $1.5$ times.
The increase in polarization degree also depends on the wavelength and disc puffing-up.
In our work for the disc without puffing-up, it is about $1.7$ times in U band and about $1.4$ times in I band.
The numerical value of the relative magnification depends on the disc model and the position of the observer.
With other parameters, we got both smaller values and slightly larger ones.
On the whole, we can conclude that the increase in the degree of polarization can easily be about $1.5$ times.
With a significant disc puffing-up, the possible increase in polarization turns out to be less significant.
Instead, the polarization plane orientation may depend on the perturbation shape:
during an eclipse by a compact cloud or a perturbation with a small extension, the disc radiation is polarized perpendicular to its plane,
and, during an eclipse by an extended perturbation, the disc radiation is polarized along its plane.

A deep minimum of UX~Ori has been observed recently by Belan and Shakhovskoy (in preparation, see also The $2^{nd}$ international Workshop \textquoteleft UX Ori type stars and related topics\textquoteright\ \href{http://uxor.ru/images/presentation/belan.pdf}{materials}).
Its depth in the V band was $\sim 3.5$~mag with the polarization degree more than $10$~per cent instead of the usual depth of $\sim 2.5$~mag and polarization degree about $6$~per cent.
\citet{Rostopchina2001} observed unusual deep minima of VV~Ser.
In the V band, its depth was $\sim 3.2$~mag, and the polarization degree was about $10.5$~per cent.
In the U band, the depth was $\sim 3.3$ mag with the polarization degree about $16$~per cent.
The eclipse was not very long, and no noticeable change in the PA was observed.
The calculations presented above show that the considered models with an extended perturbation can be applied to describe such observations.
However, the presented results significantly depend on the model.
Hence such unusual, deep and sometimes prolonged eclipses deserve a separate careful study and modelling.
This may provide new information about the shape and size of those structures in the circumstellar discs, which led to such eclipses.

In our model, changes in the dependence of the linear polarization degree on the fading level are also possible.
Near the brightness minimum, the degree of polarization may not change or even slightly decrease while the eclipse depth increases.
A similar change in the dependence of the polarization degree on the eclipse depth was observed for a classical T~Tauri star RW~Aur~A during the unprecedentedly long and deep dimming event in 2015--2016 \citep{Dodin2019}.

Thirdly, large-scale perturbations with different shapes additionally increase the parameters scatter (polarization degree and colour indices)
on the same fading level.
This is observed when considering values at the minimum brightness.
When studying the large-scale perturbation motion around the star, the effect is further enhanced.
Usually, an increase in the length of the disturbance leads to a lower degree of polarization at the same fading level,
but there may be exceptions for narrow perturbations.
Close to a deep minimum, the degree of polarization can vary by 1--2~per cent, depending on the perturbation azimuthal extension.
In general, an increase in the perturbation azimuthal extension leads to higher modulus values of colour indices.
This cause of the parameters scatter can be identified by comparing different eclipses of the same star, taking the additional analysis of the disc puffing-up into account (for example, by IR-excess).

Also, when the large-scale perturbation moves around the star, new effects arise which are not possible for an eclipse by a compact gas-dust cloud.
In a model with a significant disc puffing-up, the maximum of the polarization degree can be achieved not at the minimum brightness.
It may be slightly shifted relative to it.
In this case, at the minimum brightness, the degree of polarization can be less than a per cent,
while the maximum polarization degree will be about 3~per cent.
Observations of such eclipses with a shifted linear polarization maximum are not known to us.

The behaviour of the linear polarization PA turns out to be very complex.
For a disc without puffing-up in the dust sublimation zone, small (on the order of several degrees) changes in the position angle in both directions are possible.
As the puffing-up increases, the PA deviations become more noticeable, but at the same time, they are mainly directed in the same direction.
In some models, during an eclipse, the PA can change up to 90\textdegree.
In others, the PA at the brightness minimum does not change at all,
but its strong changes are observed after the minimum.
In both cases, these changes may present for some time after an eclipse, when the star brightness returns to pre-eclipse level.
These deviations can be up to 60\textdegree\ for a thin disc and up to 30\textdegree\ for a thick one.
The effect depends on the optical properties of the dust.
It is possible that in one spectral band the disc with a puffing-up is thin while in the other band the disc is already thick.
As a result in the first band, we have a strong change in the PA while in the second band there are smaller changes in the opposite direction.
The model with a constant mass outflow rate and the permanent perturbation shape predicts deviations of the PA before and after the eclipse that are equal in absolute value but occur in different directions.
Observations of UX~Ori and WW~Vul (Fig.~\ref{fig:UXOri}) show no such symmetry.
It may indicate both a variable mass outflow rate and a more complex form of the disc perturbation.
For example, the azimuthal extension of the perturbation may be asymmetric due to the disc rotation.

In this paper, the puffing-up is simulated by a dusty disc wind.
The wind density depends on the mass outflow rate and may change for various reasons.
Usually, the mass outflow rate is assumed to be equal to a certain fraction of the accretion rate.
For example, \cite{Bans2012} estimate the mass outflow rate as 1--5~per cent of the accretion rate.
At the same time, the accretion rate of young stars is unstable.
Its variability can reach $0.5$ dex for Herbig Ae/Be stars~\citep{Mendigutia2011, Pogodin2012}.
But a change in the wind density by $0.5$ dex could have a very strong effect on the eclipse shape.
Also, the shape of the perturbation can be quite complex and change over time.
A simultaneous change in the shape of the disc perturbation and the mass outflow rate can lead to complex changes in the stellar magnitude,
the polarization degree, and its PA in time.
This can bring us closer to understanding the nature of the unusual deep and lasting minima of UX~Ori and WW~Vul~\citep{Grininetal1988, Grininetal1994}.
At the moment we can separately obtain the observed changes in the magnitude or the PA of linear polarization,
but we cannot reproduce all the features of the eclipse at the same time.
The possible reason for such minima is the strong asymmetry of the absorbing screen.
We plan a thorough comparison of the model with observations, identification and refinement of its parameters to be carried out in the future.
In particular, consideration of a more complex perturbation shape.

We found that changes in the PA strongly depend on the dust properties and, consequently, on the wavelength of observations.
Hence, the photopolarimetric observations in different spectral bands are valuable for a deeper analysis of the phenomenon and determining the parameters of the disc and the absorbing screen.
The complex changes in the position angle can be partly explained by the variability in the rates of accretion and mass outflow, which affect the disc puffing-up.
They can be estimated from spectral observations.
Such observations during the eclipses would be essential.
In addition, significant changes in the puffing-up shape affect the SED.
Therefore, obtaining several star SEDs (for example, in the IR range) during the eclipse should also provide helpful information about these phenomena.

Thus the above conclusions of the models can be summarized as an extension of the conservative model of a compact gas-dust cloud eclipse of a star with a thin disc~\citep{Grinin1988}.
Computations for a disc with a puffed-up inner rim demonstrate that both models (the compact cloud eclipse and the large-scale perturbation eclipse)
explain several observational phenomena that do not fit the predictions of the thin disc model.
The large-scale perturbation model not only generalizes the predictions of the compact cloud model,
allowing one to obtain deeper eclipses and a greater scatter of parameters,
but also leads to new effects associated with changes in the PA of linear polarization.

A detailed study of such eclipses of UX~Ori type stars (based on various observations) can give us more information in the future about the inner regions of the protoplanetary disc and the processes occurring in them.
Now, since it is possible to obtain resolved images of the discs, the large-scale structures in the disc can be seen from the disc pole.
Our models show that, under certain conditions, large-scale disc perturbations have important observational manifestations with a line of sight close to the disc plane
and explain observed eclipses.
Since it is reasonable to expect large-scale disc structures to be independent of the disc orientation relative to an observer,
this gives us an additional way to study such structures as shadows on the disc images.
As noted by~\citet{Stolker2017}, the dense in time observations of such moving shadows can be considered as a face-on version of the studies of the UXOR phenomenon.

There is no universal dependence of polarization and colour indices on the fading level in the large-scale disc perturbation model.
Each model is different in terms of quantitative predictions.
This is very important because by modelling the results of observations, we can better understand the mechanisms of the perturbation formation in the discs.

\section*{Acknowledgements}

V.\,P.\,G.\ acknowledges the support of Ministry of Science and Higher Education of the Russian Federation under the grant 075-15-2020-780 (N13.1902.21.0039).
The authors thank the referee for useful suggestions that helped improve the manuscript.

\section*{Data Availability}

The data underlying this article will be shared on reasonable request to the corresponding author.


\bibliographystyle{mnras}
\bibliography{ShulmanGrininUXOriEclipsesByDiscPerturbations}

\appendix

\section{Multiple scatterings contribution}\label{appendix:multiple}

\begin{figure}
    \begin{center}\includegraphics[width=1\columnwidth]{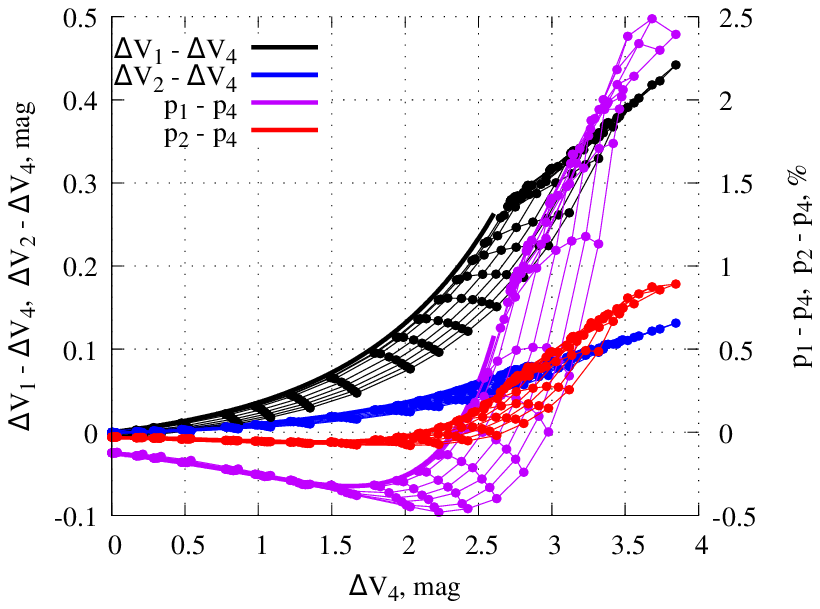}\end{center}
    \caption{The comparison of the results obtained taking into account only single ($\Delta V_1$ and $p_1$),
    first two ($\Delta V_2$ and $p_2$) and first four ($\Delta V_4$ and $p_4$) scatterings
    for the disc model from Fig.~\ref{fig:eclipse_nw} with $\Delta r = 0.3$ au and $r_0 = 0.5$ au\@.
    The angle between the direction to the observer and the disc axis is 65\textdegree.}
    \label{fig:multiple_scatterings}
\end{figure}

It is very important to take multiple scatterings into account when we study the dust radiative transfer.
Here we consider eclipses from the section~\ref{sec:res_eclipses}.
Here we use the models from Fig.~\ref{fig:eclipse_nw} with $\Delta r = 0.3$ au and $r_0 = 0.5$ au\@.
In Fig.~\ref{fig:multiple_scatterings} we show the difference between the results obtained taking into account four first scatterings
and the results obtained taking into account only single or double scatterings.
For this purpose in the current subsection we use the following notation for the main eclipse parameters:
$\Delta V_1$, $\Delta V_2$, and $\Delta V_4$ are maximal changes in the magnitude in V band  when we take into account only single, first two, and first four scatterings, respectively;
$p_1$, $p_2$, and $p_4$ are the corresponding maximal polarization degrees.
Again, we show the results for compact cloud eclipses with thick lines, and the deepest points of the large-scale humps with dots.
Thin lines connect points with equal $h_{\mathrm{hump0}}$ (isoheights) and $\Delta \phi$ (isoextensions).
Isoextensions go along the compact cloud eclipse track, and isoheights go across it.

Maximal changes in the magnitude have simple behaviour: $\Delta V_1$ is always greater than $\Delta V_2$ and $\Delta V_2$ is always greater than $\Delta V_4$.
When we take into account more scatterings, the disc becomes brighter, and the star input into the system light decreases.
As a result, the same star fading leads to a smaller magnitude change.
When the hump becomes more extended the eclipsed region of the disc grows, multiply scattered light from this region is also absorbed by the perturbation,
so the disc brightening is reduced, and the magnitude differences become smaller than in the compact cloud case.

The polarization has more difficult behaviour.
The main contribution to polarization is from the single scatterings.
After multiple scatterings, the light is only slightly polarized.
Hence, the total light of the system has three components: nonpolarized direct starlight, strongly polarized single scattered light,
and slightly polarized multiply scattered light.
As a result, when the star is weakly eclipsed, slightly polarized light increases the total polarization degree.
In this situation $p_1$ and $p_2$ are less than $p_4$.
Conversely, when the star is heavily eclipsed and makes a small contribution to the total system light,
slightly polarized multiple scattered light decreases the total polarization degree.
So, $p_1$ and $p_2$ become greater than $p_4$.

The difference between taking into account the first two or first four scatterings for our model is always less than $0.15$~mag and $1$~per cent.
The difference between the first three and four scatterings is even less (not more than $0.035$~mag and $0.3$~per cent) and we do not show it in Fig.~\ref{fig:multiple_scatterings}
so as not to overload it.

The difference between the first four and five scatterings is not more than $0.02$~mag and $0.15$~per cent.
The contribution of subsequent scattering to the total radiation of the system decreases rapidly.
Therefore, in this paper, we limited the computations to modelling only the first four scatterings.

\section{Disc wind model influence}\label{appendix:disk_wind}

\begin{figure}
    \begin{center}\includegraphics[width=1\columnwidth]{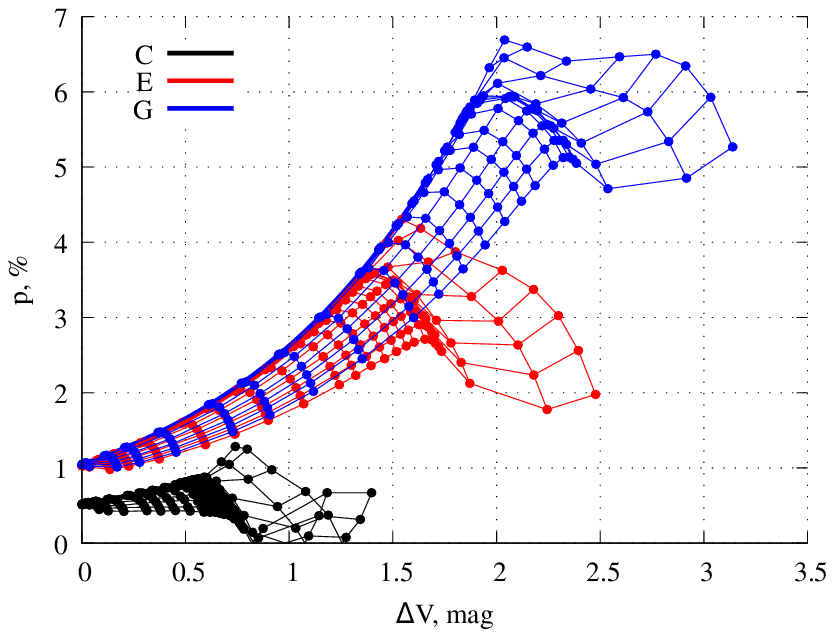}\end{center}
    \caption{The same as in Fig.~\ref{fig:wind_mout} for the disc with the disc wind.
    The mass outflow rate is $10^{-8}$ $M_\odot$ year$^{-1}$.
    Wind models C, G, and E are used.}
    \label{fig:wind_model}
\end{figure}

For wind model comparison we selected three models: C, E, and G\@.
C model gives the strongest puffing-up among all models, model G~--- the weakest, and model E is one of the intermediate.
These parameters for three wind models used in our paper are listed in table~\ref{tab:wind_models}.

\begin{table}
    \centering
    \caption{Parameters of the disc wind models.}
    \label{tab:wind_models}
    \begin{tabular}{lccc}
        \hline
        Model  & $\kappa_w$    & $\lambda_w$   & $\xi_0'$  \\
        \hline
        C       & $0.01$        & $75.43$       & $1.73$    \\
        E       & $0.10$        & $25.63$       & $3.73$    \\
        G       & $0.01$        & $189.34$      & $3.73$   \\
        \hline
    \end{tabular}
\end{table}

The model C disc wind rises from the disc at a large angle to the disc plane compared to models E and G\@.
As a result, the disc puffing-up with model C wind is more noticeable.
The mass loading in model E is higher than in model G, so the wind model G provides the less significant puffing-up among the three considered models.

In Fig.~\ref{fig:wind_model} we show the results for the same observer position and the parameters of the humps as in Fig.~\ref{fig:wind_mout}.
But this time the mass outflow rate is always $10^{-8}$ $M_\odot$ year$^{-1}$, and the wind models are $C$, $G$, and $E$.

One can see that the wind model choice may give a difference similar to changing mass outflow rate several times.
So, it is rather tricky to estimate the mass outflow rate based on the eclipse shape.
At the same time, we should conclude that the wind may strongly influence the eclipse and lead to new phenomena.
The disc wind is only one of possible reasons for the disc puffing-up in the dust sublimation zone.
The conclusions above should be generally true for any puffing-up in this zone regardless of its physical nature.

\bsp	
\label{lastpage}
\end{document}